# THE RELATION BETWEEN DISENTANGLED STATES AND ALGORITHMIC INFORMATION THEORY

Rubens Viana Ramos

rubens@deti.ufc.br

*Department of Teleinformatic Engineering – Federal University of Ceara - DETI/UFC*

*C.P. 6007 – Campus do Pici - 60755-640 Fortaleza-Ce Brazil*

In this short communication it is proposed the general form of a n-qubit disentangled state as a irreducible sentence, in the sense explained by the algorithmic information theory, whose length increases in a non-polynomial way when the number of qubits increases.

Quantum entanglement measure is a crucial point in quantum information theory. Several entanglement measures have been proposed [1-4]. Among these, the quantum relative entropy is of particular interest. It is based on the distance between the state whose entanglement one wishes to measure and the set of disentangled states. In order to calculate it, it is interesting to know the general form of an *n*-way disentangled state. It is obvious that, if a *n*-partite of qubit quantum state cannot be written in the form of a general *n*-partite disentangled stated, then it is entangled. For example, a bipartite state $\Gamma_{AB}$ has 2-way entanglement if

$$\Gamma_{AB} \neq \sum_i p_i \left( \rho_A^i \otimes \rho_B^i \right) \tag{1}$$

$$\sum_i p_i = 1 \tag{2}$$

where $\rho_A$ and $\rho_B$ are single-qubit states. On the other hand, a tripartite state $\Gamma_{ABC}$ has 3-way entanglement if

$$\Gamma_{ABC} \neq \sum_i p_i \left( \rho_A^i \otimes \rho_B^i \otimes \rho_C^i \right) + \sum_j r_j \left( \rho_A^j \otimes \Phi_{BC}^j \right) + \sum_l q_l \left( \rho_B^l \otimes \Phi_{AC}^l \right) + \sum_k t_k \left( \rho_C^k \otimes \Phi_{AB}^k \right) \tag{3}$$

$$\sum_i p_i + \sum_j r_j + \sum_l q_l + \sum_k t_k = 1 \tag{4}$$

where $\Phi_{AB}$, $\Phi_{BC}$ and $\Phi_{AC}$ are entangled bipartite states. When the number of qubits $n$ increases, the number of terms of the general $n$-way disentangled state grows very fast. For 5 qubits it has 66 terms, for 6 qubits it has 332 terms while for 7 qubits it has 1681 terms. Since any term of the general $n$-way disentangled state represents physically a different kind of quantum state in the meaning that the positions of the entanglement are different, the sentence cannot be reduced to a smaller one containing a lower number of terms. For a tripartite state, for example,

$$\rho_A^i \otimes \rho_B^i \otimes \rho_C^i \neq \sum_j r_j \left( \rho_A^j \otimes \Phi_{BC}^j \right) + \sum_l q_l \left( \rho_B^l \otimes \Phi_{AC}^l \right) + \sum_k t_k \left( \rho_C^k \otimes \Phi_{AB}^k \right) \tag{5}$$

$$\rho_A^j \otimes \Phi_{BC}^j \neq \sum_i p_i \left( \rho_A^i \otimes \rho_B^i \otimes \rho_C^i \right) + \sum_l q_l \left( \rho_B^l \otimes \Phi_{AC}^l \right) + \sum_k t_k \left( \rho_C^k \otimes \Phi_{AB}^k \right) \tag{6}$$

Thus, for (6), for instance, one cannot construct a 3-way disentangled state having 2-way entanglement between subsystems $B$ and $C$, mixing completely disentangled tripartite states and tripartite states having 2-way entanglement between systems $A$ and $C$, and $A$ and $B$. Hence, this work proposes the general $n$-way disentangled state as an irreducible sentence, in the same sense that $\Omega$ [5,6] proposed by Chaitin is an irreducible binary number. Since the number of terms of a general $n$-qubit disentangled state grows in a non-polynomial way, any problem where it is necessary and the number of qubits increases is a NP problem that can not be reduced to a P problem. This means that entanglement measures based on the search of minimum distance using the general disentangled state can not be calculated in an efficient way when the number of qubits increases.